# Ponderomotive scattering of an electron-bunch before injection into a laser wakefield

A. G. Khachatryan, M. J. H. Luttikhof, F. A. van Goor, and K.-J. Boller

Faculty of Science and Technology, University of Twente, P.O. Box 217, 7500 AE Enschede, The Netherlands

For the purpose of laser wakefield acceleration, it turned out that also the injection of electron bunches longer than a plasma wavelength can generate accelerated femtosecond bunches with relatively low energy spread. This is of high interest because such injecting bunches can be provided, e.g., by state-of-the-art photo cathode RF guns. Here we point out that when an e-bunch is injected in the wakefield it is important to take into account the ponderomotive scattering of the injecting bunch by the laser pulse in the vacuum region located in front of the plasma. At low energies of the injected bunch this scattering results in a significant drop of the collection efficiency. Larger collection efficiency can by reached with lower intensity laser pulses and relatively high injection energies. We also estimate the minimum trapping energy for the injected electrons and the length of the trapped bunch.

In Laser Wakefield Accelerator (LWFA) a femtosecond high-power laser pulse generates a strong plasma wave (laser wakefield) which can accelerate charged particles to ultra-relativistic energies [1]. In order to avoid large energy spread in accelerated bunch when a relativistic electron bunch is injected in the laser wakefield, the bunch has to be injected at a suitable position in the wake with a precision of a fraction of the plasma wave period and the duration of the injected bunch has to be similarly short. For the plasma parameters of interest, which involves typical plasma wavelength of a few tens of microns, this requires initial bunches of the order of 10 femtoseconds duration and a similar precision of the synchronization with the wakefield. With current technologies these requirements cannot be fulfilled in practice. As an alternative, in the recently demonstrated "bubble" injection method [2] electrons from the background plasma are trapped in the correct phase of the wake yielding the required ultra-short bunches. This method led to acceleration to energies of the order of 100 MeV with relatively low energy spread. However, the shot-to-shot reproducibility was reported to be poor so far. Correspondingly, other injection schemes have been proposed aiming on a better control of the experimental conditions. Among these controlled injection methods are all-optical methods [3], when electrons from the plasma itself can be trapped and accelerated by the use of additional laser pulse(s).

Recently it turned out that a relatively long (longer than plasma wavelength) external electron-bunch can be used in LWFA as an injected bunch [4-8]. Simulations showed that, under certain conditions, electrons from the bunch can be trapped in a small region near the axis. The trapped bunch is accelerated in the wakefield, and this should result in an extremely short output bunches with a few microns in sizes, with high energy (hundreds of MeV's), and with relatively low energy spread (in the order of one or a few percents). The injecting few-MeV's e-bunch can be obtained from state-of-the-art RF photo injectors. This determines high interest in the external e-bunch injection schemes for LWFA.

In this paper we point out that when an electron bunch is injected in the wakefield just *behind* the laser pulse [7,8] one has to take into account the following fact. Because the electron bunch propagates slower than the laser pulse in vacuum, at some distance before the plasma (actually in vacuum) the bunch becomes situated in *front* of the laser pulse (this situation is schematically depicted in Fig. 1). Therefore the bunch is exposed to the laser pulse while traveling through vacuum and one has to take into consideration the interaction of the injecting bunch and the pulse before both enter the plasma. The distance from the plasma at which the bunch and the pulse start to overlap, $L_c$ (catching distance), can be calculated from the following expression: $L_c=(l_b+l_L)/(1-\beta_b)$, where $l_b$ ($l_L$) is the length of the e-bunch (laser pulse), $\beta_b=v_b/c$ is the bunch velocity normalized to the speed of light. When $\gamma_0^2=1/(1-\beta_b^2)>>1$ one has $L_c \approx 2\gamma_0^2(l_b+l_L)$, here $\gamma_0$ is the relativistic factor of the injecting bunch. For example, when the kinetic energy of the initial bunch is 1.6 MeV ($\gamma_0 \approx 4.13$) and $l_b+l_L \approx 180$ μm (parameters from Fig. 5 in Ref. [7]), the catching distance is ≈6 mm. For comparison, the Rayleigh length, $Z_R=\pi w_0^2/\lambda_L$, is ≈3.5 mm for a laser spot radius $w_0$=30 μm and laser wavelength $\lambda_L$=0.8 μm, which are typical values for simulations and experiments. So, in this example, when the bunch passes through the pulse, the laser intensity is comparable with its value in the focus. Thus, at sufficiently high intensity in the focus, the ponderomotive force [9] might lead to strong scattering of the bunch before injection into the wakefield. In this paper we are aimed to take this effect into account. It is also clear that at higher initial energies of the e-bunch, the major part of the bunch interacts with the laser pulse at relatively large distance from the focus (the entrance of the plasma) where the ponderomotive force is weak, so that the effect of the ponderomotive scattering could be small.

First, let us estimate such important parameters of the problem as the minimum trapping energy for injecting electrons and the length of a trapped bunch, which determines the relative energy spread in an accelerated bunch. Suppose that a mono-energetic e-beam is injected in all phases of the wakefield. Following Ref. [7], we assume that the on-axis wake potential is $\phi=\phi_m \sin(\psi)$, where $\psi=k_p z-\omega_p t$, $k_p$ ($\omega_p$) is the plasma wavenumber (frequency), and consider the integral of motion (Hamiltonian) for injected electrons,

$$\Gamma(\gamma) - \phi_m \sin(\psi) = const, \qquad (1)$$

where $\Gamma=\gamma[1-\beta_g(1-\gamma^{-2})^{1/2}]$, $\gamma$ is the relativistic factor of an electron, $\beta_g=v_g/c$ is the normalized group velocity of the laser pulse. Fig. 2 shows typical behavior of the wakefield components near the axis, where, under certain conditions, part of the injected bunch is trapped. The phase region suitable for trapping and acceleration of electrons, where the wakefield is accelerating and focusing, is shown by two dotted vertical lines and is given by $\psi_- \leq \psi \leq \pi/2$. In a uniform plasma $\psi_-=0$, however in a plasma channel the focusing region becomes broadened, so that the minimum trapping phase is shifted somewhat to the left [6,7], and $\psi_-<0$. Suppose that an electron with a relativistic factor $\gamma_0<\gamma_g=(1-\beta_g^2)^{-1/2}$ is injected at $\psi=\psi_0$. The injected electron slips backwards relative to the laser pulse and can gain energy in the accelerating region. If the wake amplitude is sufficiently high and the electron is injected at a proper phase, it can reach $\gamma=\gamma_g$ at the trapping (return) point, and then accelerated to an ultra-relativistic energy. The trapping phase, $\psi_t$, can be found from the Hamiltonian (1):

$$\sin(\psi_t) = -q + \sin(\psi_0), \qquad (2)$$

where $q=[\Gamma(\gamma_0)-\Gamma(\gamma_g)]/\phi_m$. Taking into account that $\psi_- \leq \psi_t \leq \pi/2$ and $q \geq 0$, one has the following trapping condition for the initial phases:

$$\sin(\psi_0) \geq q + \sin(\psi_-) \equiv S. \qquad (3)$$

Obviously, when $S>1$ no electrons can be trapped in the wakefield because the initial energy of the injected beam is too low for the chosen wakefield. For a higher injection energy, when $S=1$, only electrons injected at $\psi_0=\pi/2$ can be trapped, and the trapping point is at $\psi_t=\psi_-$. The minimum relativistic factor for trapping, $\gamma_{min}$, can be found from $S=1$. When $\gamma_g^2>>\gamma_0^2>>1$ one has the following expression:

$$\gamma_{\min} \approx \frac{1}{2[(1-\sin(\psi_-))\phi_m + \gamma_g^{-1}]}, \qquad (4)$$

which is in a good agreement with numerical simulations. When $\gamma_0 > \gamma_{\min}$ ($S<1$) the region of initial phases for trapped electrons becomes $\pi/2 - \delta\psi_0/2 \leq \psi_0 \leq \pi/2 + \delta\psi_0/2$, where $\delta\psi_0$ is the width of the region (we assume that the wakefield is focusing within this region). This region is shown in Fig. 2 by the transparent rectangular. The trapped electrons occupy the phase region $\psi_- \leq \psi_t \leq \psi_- + \delta\psi_t$, depicted by the black rectangular in Fig. 2. One can show that electrons at the borders of the initial trapping region, with $\psi_0 = \pi/2 \pm \delta\psi_0/2$, are trapped at the same point, namely at $\psi_t = \psi_-$, and electrons with $\psi_0 = \pi/2$ are trapped at $\psi_t = \psi_- + \delta\psi_t$. The width of the initial region for trapped electrons, $\delta\psi_0$, is determined by the equality $\sin(\psi_0) = S$ [7], from which one finds $\delta\psi_0 = 2\arccos(S)$. The collection efficiency, that is the ratio of number of trapped electrons to the total number of injected particles, is $\delta\psi_0/2\pi$ [7]. The width of the trapping region, $\delta\psi_t$, can be calculated taking into account that $\psi_t = \psi_- + \delta\psi_t$ corresponds to $\psi_0 = \pi/2$. In this case, from Eq. (2), we have $\delta\psi_t = \arcsin(1-q) - \psi_-$. This gives for the trapped bunch length:

$$l_t = (\lambda_p/2\pi)[\arcsin(1-q) - \psi_-], \qquad (5)$$

where $\lambda_p$ is the plasma wavelength. In the most interesting case of a short trapped bunch ($\delta\psi_t << 1$) one has $\delta\psi_t \approx [\Gamma(\gamma_{\min}) - \Gamma(\gamma_0)]/\phi_m \cos(\psi_-)$, and if additionally $\gamma_g^2 >> (\gamma_{0,\min})^2 >> 1$, $\delta\psi_t \approx (1/\gamma_{\min} - 1/\gamma_0)/2\phi_m \cos(\psi_-)$. For example, when $\psi_- = -0.75$, $\lambda_p = 47$ μm, $\gamma_g = 59$, $\phi_m = 0.1$ and the injection energy is 1.6 MeV ($\gamma_0 \approx 4.13$), expression (5) predicts the trapped bunch length of $\approx 5$ μm, that agrees with the simulations (see Fig. 5(b) in [7]). So, when the injection energy is close to the minimum trapping energy, the trapped bunch length can be much smaller than the plasma wavelength. In this case the trapped electrons experience almost the same accelerating force that could lead to a small energy spread in the accelerated bunch [7]. The higher the initial bunch energy, the larger the collection efficiency and the trapped bunch length, but also the larger the energy spread. Thus, the above estimations predict that a long (unphased) e-bunch injection in the laser wakefield could lead to a very short accelerated bunches with a relatively small energy spread.

To study the effect of interaction of the injected bunch with the laser pulse before the plasma channel, we describe the axially-symmetrical Gaussian laser pulse with the normalized amplitude, $a = E_L/(m_e c \omega_L/e)$ [1], as $a = a_0(w_0/w)\exp[-r^2/w^2 - (\psi-\psi_c)^2/\sigma^2]$, where $\psi_c$ is the center of the pulse. The pulse is focused and matched to a plasma channel and guided in it with a constant radius. In the channel $w=w_0$ and $w=w_0(1+z^2/Z_R^2)^{1/2}$ before the channel, in vacuum, where the ponderomotive force [9] is used to calculate the bunch-laser interaction. The injected bunch is modeled by a random Gaussian distribution in both longitudinal and radial direction, with an average electron concentration $n_b \sim \exp[-r^2/r_b^2-(z-v_bt)^2/\sigma_b^2]$. The channel-guided laser wakefield and electron motion in the wakefield are calculated the same way as done in [5].

To demonstrate the effect of ponderomotive scattering of the injected bunch in front of the channel, we first choose parameters for the wakefield and the injected bunch energy used for Fig. 5 of Ref. [7]. Our Fig. 3 shows the injected bunch at the entrance of the plasma channel without (a) and with (b) the laser-bunch interaction before the channel. One can see strong ponderomotive scattering of most of the injected electrons in Fig. 3(b); the rms spread in the transverse momentum in this case is about 0.1 $m_ec$. As a result the collection efficiency becomes very small: it is 1 % (26 %) with (without) inclusion of the interaction.

For our simulations we choose the parameters close to those used in [7]: $\lambda_L$=0.8 μm, $\lambda_p$=47 μm (plasma electron concentration of ≈5×10$^{17}$ cm$^{-3}$), $w_0$=30 μm, $\sigma$=15 μm, the spot-size-corrected gamma factor, corresponding to the laser group velocity in the channel of $(\lambda_p/\lambda_L)[1+(\lambda_p/\pi w_0)^2]^{-1/2}$≈52.6. The e-bunch is focused to the channel to a radius $r_b=w_0$ and with a convergence angle of 0.42°, so that at 20 cm before the channel the bunch radius is 1.5 mm. The full-width-at-half-maximum duration of the injected bunch is 250 fs (that corresponds to $\sigma_b$=45 μm), the bunch center is at 60 μm distance from the laser pulse center when the bunch enters the channel. The plasma channel length is fixed and equal to 5 cm. For these parameters of the problem four trapped bunches are formed. The values are calculated for *all* trapped electrons. Fig. 4 compares the collection efficiency, the relative energy spread and the mean energy in the accelerated bunch with and without the interaction in front of the channel in dependence on the injection energy (in this case $\phi_m$=0.1). Again we see that when one takes into account the ponderomotive scattering, the collection efficiency drops considerably at low injection energies. At higher injection energies the effect of the scattering becomes weaker, as mentioned above, however the energy spread in the accelerated bunch in this case becomes larger. In Fig. 4 some improvement in the relative energy spread can be seen with the interaction turned on, due to smaller contribution from the off-axis electrons. In Fig. 5 we plot the collection efficiency and the relative energy spread in the accelerated bunches for the case of lower laser intensity, $a_0$=0.2, for which $\phi_m$≈0.016 and the minimum trapping energy is ≈5.5 MeV ($\gamma_0$≈12). Simulations showed that in this example the effect of ponderomotive scattering is negligible because of the lower pulse intensity and larger catching distance (remember that $L_c \propto \gamma_0^2$), so that most of the injected electrons interact with the laser pulse at large distance from the focus, where the ponderomotive force is very weak. For the injection energies used for Fig. 5, the mean energy in the accelerated bunch lies between 40 and 49 MeV's.

In summary, we studied an important additional issue that may have significant influence on external e-bunch injection for laser wakefield acceleration. It turns out that, due to the laser-bunch interaction in front of the plasma, a high collection efficiency together with a small energy spread of the accelerated e-bunch can be obtained only with a combination of relatively high injection energy and a weak laser wakefield. The energy gain in this case is typically in the order of a few tens of MeV's. Such an experiment is planned for the near future at Eindhoven University of Technology [10].

This work is supported by the Dutch Foundation for Fundamental Research on Matter (FOM) within the "Laser Wakefield Accelerator" program.

FIGURE CAPTIONS

FIG. 1. Schematic view of electron bunch injection in the channel-guided laser wakefield. The bunch and the pulse are focused into the channel.

FIG. 2. The wake potential, $\phi$, longitudinal, $E_z$, and transverse, $f_r$, components of the wakefield vs. the injection phase. The dotted vertical lines show the trapping region ($E_z<0$, $f_r<0$). The black rectangular represents the trapped bunch, and the transparent rectangular shows the initial phase range for the trapped electrons.

FIG. 3. (Color) The electron concentration (in arbitrary units) in the injected bunch, at the entrance of a plasma channel, without (a) and with (b) ponderomotive scattering by the laser pulse in front of the channel. The initial kinetic energy of the bunch is 1.6 MeV, the normalized laser pulse amplitude at focus is $a_0$=0.5, the wake amplitude is $\phi_m$=0.1. The circles depict the position of the laser pulse.

FIG. 4. The collection efficiency, relative energy spread and the mean energy in the accelerated bunch vs. the kinetic energy of the injected bunch. The lines 1 (2) show the values obtained without (with) the bunch scattering by the laser pulse in front of the channel. $a_0$=0.5 (corresponding peak intensity is $I_0 \approx 5.3 \times 10^{17}$ W/cm$^2$), $\phi_m$=0.1, other parameters are given in the text.

FIG. 5. The collection efficiency (1) and the relative energy spread (2) in the case of a lower intensity laser pulse, $a_0$=0.2 ($I_0 \approx 8.4 \times 10^{16}$ W/cm$^2$). Other parameters are the same as in Fig. 4.

FIG. 1

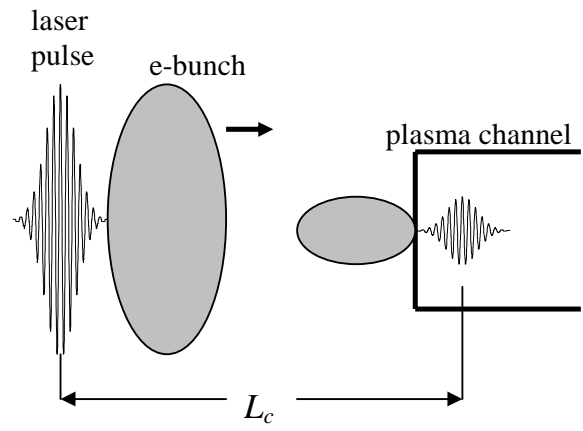

FIG. 2

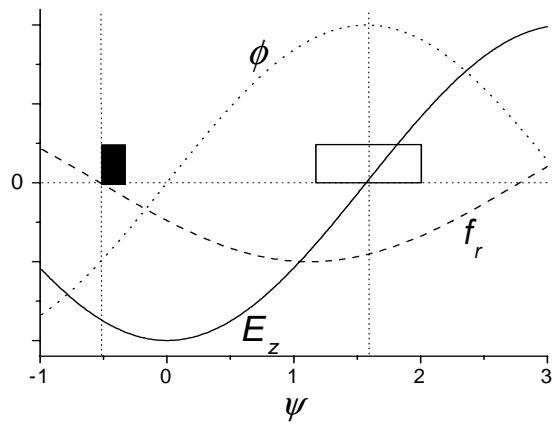

FIG. 3

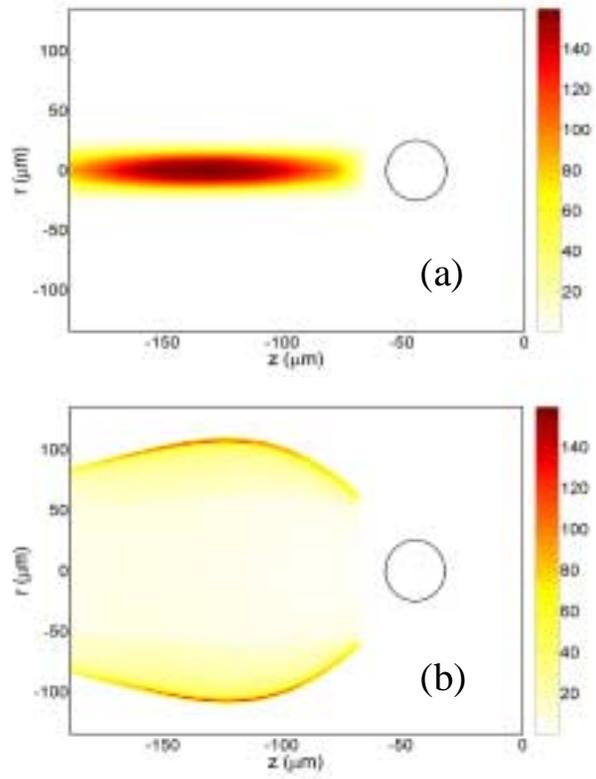

FIG. 4

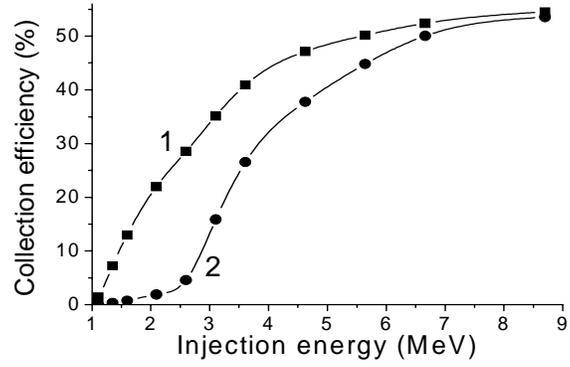
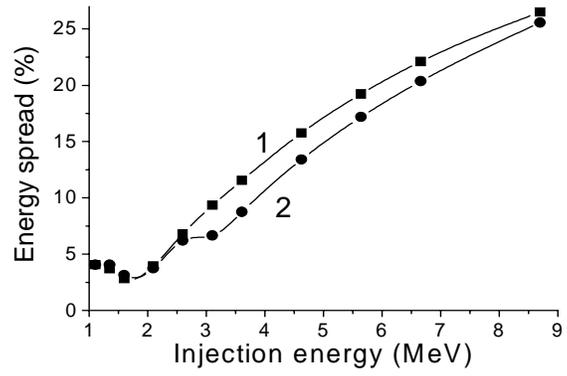
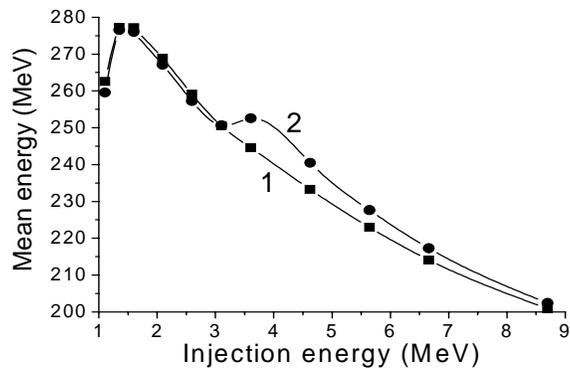

FIG. 5

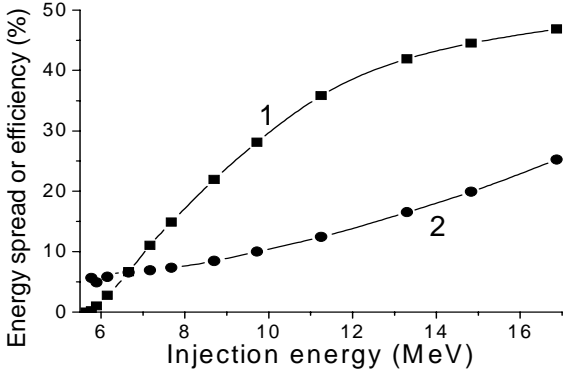